\UseRawInputEncoding
\documentclass{aastex631}

\usepackage{aas_macros}

\shorttitle{Extended DE model with massive neutrinos}
\shortauthors{Sharma et al.}
\graphicspath{{./}{figures/}}

\usepackage{graphicx}	
\usepackage{amsmath}	
\usepackage{amssymb}	
\usepackage{booktabs}

\def\beq{\begin{equation}}
\def\eeq{\end{equation}}
\def\bea{\begin{eqnarray}}
\def\eea{\end{eqnarray}}

\begin{document}

\title{Implications of an extended dark energy model with massive neutrinos}

\author{Ravi Kumar Sharma}
\affiliation{Indian Institute of Astrophysics, Bengaluru, Karnataka 560034, India}
\email{ravi.sharma@iiap.res.in}

\author{Kanhaiya Lal Pandey}
\affiliation{Indian Institute of Astrophysics, Bengaluru, Karnataka 560034, India}
\email{kanhaiya.pandey@iiap.res.in}

\author{Subinoy Das}
\affiliation{Indian Institute of Astrophysics, Bengaluru, Karnataka 560034, India}
\email{subinoy@iiap.res.in}




\begin{abstract}
Recently there has been reports of finding a lower bound on the neutrino mass parameter ($\Sigma m_\nu$) when using ACT and SPTpol data however these bounds on the $\Sigma m_\nu$ are still weaker for most case around at 1 $\sigma$ level. In this context, here in this work, we study the consequences of using an enlarged four parameter dynamical dark energy equation of state on neutrino mass parameter as well as on the Hubble and S8 tensions. The four parameter dark energy equation of state incorporates a generic non-linear monotonic evolution of the dark energy equation of state, where the four parameters are the early and the present value of the equation of state, the transition scale factor and the sharpness of the transition. We report that with lensing-marginalized Planck + BAO + Pantheon and prior on absolute magnitude $M_B$  and KIDS/Viking $S_8$ prior, the model favours a non-zero value for the neutrino mass parameter at the most at ∼ 1 $\sigma$ level 
($\Sigma m_\nu = 0.1847_{-0.165}^{+0.0698}$ eV).
In this case this model also brings down the Hubble tension to ∼ 2.5~$\sigma$ level and the S8 tension to $\sim$ 1.5~$\sigma$ level. This model also provide a tighter constraints on the value of the dark energy equation of state at present epoch $w_0$ 
($w_0 = -0.9901_{-0.0766}^{+0.0561}$)
in comparison to the CPL like parameterization.

\end{abstract}




\section{Introduction}

The observations of type Ia supernovae show that the expansion of the Universe is accelerating. The acceleration requires the Universe to be dominated by an exotic fluid with negative pressure. The simplest explanation for dark energy is the cosmological constant or vacuum energy that explains the acceleration of the Universe. Though the cosmological constant is preferred from cosmological observations yet its theoretical understanding has been questionable \cite{Sahni:2002kh}. The other alternatives of the cosmological constant that can act as dark energy are scalar fields such as quintessence field, modified gravity, phantom dark energy \cite{2016JCAP...05..067B,Nojiri:2017ncd,2021PhRvD.103h3523F,2021PhRvD.103d3528B}. There has  not been any observational evidence of such alternatives but it has not been ruled out either. One such model is dynamical dark energy driven by a slowly rolling scalar field. If this is true, then it opens up many new observational windows which may shed light on the fundamental nature of this mysterious component of the Universe. 

The other reason to explore beyond the $\Lambda$CDM model is the recently emerging and persistent anomalies in present high precision cosmological data. The mismatch between values of Hubble parameter inferred from CMB data and direct measurements is one of them. The SH0ES (Supernovae H0 for Equation of State) team has measured the value of Hubble parameter $H_0=73.2\pm1.3$ km/s/Mpc using the distance ladder method \cite{Riess:2019cxk,Riess:2020fzl}. However The Planck 2018 measurement of CMB (Cosmic Microwave Background) has measured the value of Hubble parameter $H_0=67.36\pm0.54$ km/s/Mpc using $\Lambda$CDM model \cite{2020A&A...641A...6P}. So there is a $4.2\sigma$ discrepancy between both measurements. This mismatch gained significance with various improved precision measurements see, \cite{Freedman:2017yms,Schoneberg:2021qvd,DiValentino:2020vnx,DiValentino:2021izs,DiValentino:2020zio}.

Similarly there is another tension related to the measured value of  $S_8\equiv\sigma_8(\Omega_m/0.3)^{0.5}$, where $\sigma_8$ is the root mean square of matter fluctuations on a 8 $h^{-1}$Mpc scale,
and $\Omega_m$ is the total matter abundance. The latest prediction from Planck CMB data within the $\Lambda$CDM framework is $S_8=0.832 \pm0.013$ \cite{2020A&A...641A...6P}.

Originally, observations of galaxies through weak lensing by the CFHTLenS collaboration have indicated that the $\Lambda$CDM model predicts a $S_8$ value that is larger than the direct measurement at the $2\sigma$ level \cite{Heymans:2013fya,MacCrann:2014wfa}.  
This tension has since then been further established within the KiDS/Viking data \cite{Hildebrandt:2018yau,Joudaki:2019pmv}, but is milder within the DES data \cite{Abbott:2017wau}. 
However, a re-analysis of the DES data, combined with KiDS/Viking, leads to a determination of $S_8$ that is discrepant with Planck at the $3\sigma$ level, $S_8=0.755_{-0.021}^{+0.019}$  \cite{Joudaki:2019pmv}. 
Recently, the combination of KiDS/Viking and SDSS data has established $S_8=0.766^{+0.02}_{-0.014}$ \cite{Heymans:2020gsg}. However, a study in \cite{Nunes:2021ipq} shows fainter $S_8$ tension when redshift-space distortions (RSD) data is inlcuded.

There has been a wide range  of solutions proposed to solve these cosmological tensions which requires new physics/modifications in the early Universe i.e. pre-recombination era  as well as in the late Universe. Not a single model yet fully solves  both $H_0$ and $S_8$ tensions simultaneously.
The class of solutions which  invokes modifications of the late-time Universe dynamics in dark energy generally  leaves $r_s$ unaffected by construction and has been studied extensively in recent times. 
The higher value of $H_0$ is then accommodated by a smaller value of $\Omega_{\rm DE}$ or $\Omega_m$ at redshift below $z_*$ such that $d_A(z_*)$ stays unaffected as well. This can be done for instance by invoking variations in the dark-energy equation of state \cite{2018arXiv181104083P,2019ApJ...876..143B,2018PhRvD..97l3504P,2017PhRvD..96b3523D,2016PhLB..761..242D,2015PhRvD..92l1302D,Poulin:2018cxd,Gogoi:2020qif,Visinelli:2019qqu,Vagnozzi:2019ezj,Hazra:2022rdl} or decaying dark matter \cite{2020JCAP...07..026P} \cite{Poulin:2016nat} \cite{Abellan:2021bpx} or non-thermal dark matter \cite{Das:2021pof} with certain level of success. 

The study of the dynamical behavior of dark energy is often done in terms of its equation of state $w(z)=\frac{P(z)}{\rho(z)}$ that can vary as a function of redshift. Equation of state $w=-1$ corresponds to the cosmological constant. 
There are some recent studies where it has been shown that solving of $H_0$ and $S_8$ tensions require $w(z)<-1$ at some $z>0$ and time-varying dark energy equation of state which cross the phantom barrier \cite{2022arXiv220201202H}. Also it has been shown that a large class of quintessence ($w > -1$) models including the ones which arise from string swampland conjecture lower the $H_0$ parameter and thereby makes $H_0$ tension worse \cite{2021PhRvD.103h1305B}, however, interacting DE models, for example \cite{Das:2005yj} where the variation in $w(z)$ starts from a higher redshift might indeed alleviate the $H_0$ anomaly.
From observations, it's required that equations of state at present time should be consistent with value $w\approx-1$, however, constraints on the equation of state at higher redshift are weaker. There have  already been several efforts to parameterize the equation of state of dark energy. Some recent works in this direction can be found in \cite{PhysRevD.97.043503,Martins:2018bzo,Marcondes:2017vjw,Yang:2021flj,Colgain:2021pmf,Chevallier:2000qy,Linder:2002et,Jassal:2005qc,Efstathiou:1999tm,Barboza:2008rh,Li:2019yem,Yang:2021eud,Jaber:2017bpx,Roy:2022fif,Heisenberg:2022lob,Theodoropoulos:2021hkk,Mawas:2021tdx,Anchordoqui:2021gji,2012Ap&SS.342..155B}. 

We explore in detail the possibility of dynamical DE with a more general model-independent approach where we go beyond the 
 CPL  (Chevalier-Polarski and Linder) parameterization \cite{2003PhRvL..90i1301L,2001IJMPD..10..213C}  where the dark energy equation of state $w$ evolves linearly with expansion factor $a$.
To be specific, in this paper, we study a generic  non-linearly evolving equation of state.
Some of the recent works on dynamical dark energy scenario such as \cite{Colgain:2021pmf} suggest that CPL parameterization is not sensitive at low redshifts and thus provide motivation for going beyond CPL like parameterization.
It was recently pointed out that if late-time cosmology is modified through time-varying w, one indeed should use the direct prior on the absolute magnitude of supernovae, $M_B$ instead of $H_0$ prior \cite{2019MNRAS.483.4803L,2020PhRvD.101j3517B,Camarena:2021jlr,2021MNRAS.505.3866E}. To our knowledge, this work is the first work where we present a detailed analysis of a four parameter dynamical DE model.
To do so, we use a generic four parameter model of dynamical dark energy equation of state $w$ originally proposed in \cite{2003PhRvD..67f3521C} and test it against the recent Planck-2018, Pantheon and BAO datasets. In comparison to CPL parameterization, this parameterization has two extra parameters to incorporate the possible non-linear evolution of the equation of state with time. The main interest of this parameterization is that it captures possible transition in the equation of state of the dynamical dark energy during the course of its evolution, which many quintessence/K-essence and phantom dark energy models exhibit \cite{2003PhRvD..67f3521C}. 

Earlier studies have shown that the total neutrino mass parameter $\Sigma m_\nu$ show significant amount of degeneracy with other cosmological parameters (eg., with the matter density parameter $\Omega_m$ and the Hubble parameter $H_0$) if we just use the CMB data alone. Moreover models with a variable $w(z)$, the constraint from CMB is essentially on one number that is the effective equation of state $w_{\rm eff}$ \citep{2010MNRAS.405.2639J}. Adding a low redshift measurement data such as the BAO and the SNe data can help in breaking these degeneracies \citep{2016PhRvD..93h3527D,2018MNRAS.477.1913S} and thus getting a better constraints on the $\Sigma m_\nu$ and the DE parameters.

In this study , we find that all four parameters of the equation of state can not be constrained fully with current observational data. Especially, Planck 2018 data alone has  poor constraining ability on dark energy parameters. Once we include the BAO and Pantheon data, the constraints improve  and the Hubble tension comes down to $2.5\sigma$ level from SH0ES measurement and $S_8$ tension comes down to $1.5\sigma$ from KIDS/Viking measurement.

An important aspect of this paper is to get neutrino mass constraints in the 4pDE model. Standard massive neutrinos play an important role in the evolution of the Universe, they leave a non-negligible impact on the cosmic microwave background (CMB) and large-scale structure (LSS) at different epochs of the evolution of the Universe. This impact is used to get a bound on neutrino mass. Some of the effects of standard model neutrinos and dark energy are the same during specific cosmic time. Therefore nature of dark energy has an important role in constraining neutrino mass. Some of the relevant studies we find in the literature are \cite{PhysRevLett.95.221301} \cite{PhysRevD.96.043510} \cite{PhysRevD.83.123504} \cite{DiValentino:2021imh} \cite{2018PhRvD..97l3504P} \cite{Vagnozzi:2018jhn} \cite{Vagnozzi:2017ovm} \cite{Abellan:2021rfq}. In our analysis, we detect a non-zero neutrino mass at $1 \sigma$ level ($\Sigma m_\nu \sim 0.2 \pm 0.1$ eV) but consistent with zero at 2$\sigma$ level unlike a previous study \cite{2018PhRvD..97l3504P} where the analysis was done with earlier (2015) Planck data and the  neutrino mass $\Sigma m_\nu$ was found to be  non-zero even at $\gtrsim 2 \sigma$.

The plan of the paper is as follows. A brief description of the four parameter dynamical dark energy equation of state, $w_{\rm de}(a)$, is given in  section II. In Section III and IV we provide a detailed description of our analysis and results. Then Section V summarizes the paper and future outlook.

\section{Four-parameter Model for Dark Energy}
\label{sec:model}

To investigate the effect of a non-linearly evolving dark energy equation of state, we use a model independent, 4 parameter dynamical dark energy equation of state $w_{\rm de}(a)$, suggested by \cite{2003PhRvD..67f3521C},
\beq
w_{\rm de}(a) = w_0 + (w_m-w_0) \times \Gamma(a)
\label{eq:wde_1}
\eeq
where $w_0$ and $w_m$ are 2 parameters denoting the initial and final values of the dark energy equation of state, 
ie., $w_0 = w_{\rm de}(a=1)$ and $w_m = w_{\rm de}(a \ll 1)$. The factor $\Gamma(a)$ contains the other 2 parameters describing the course of the evolution of $w_{\rm de}(a)$ 
, and is given as, 
\beq
\Gamma(a) = \frac{1-\exp \left(-({a-1})/{\Delta_{\rm de}}\right)}{1-\exp \left({1}/{\Delta_{\rm de}}\right)} \times \frac{1+\exp \left({a_{\rm t}}/{\Delta_{\rm de}}\right)}{1+\exp \left(-({a-a_{\rm t}})/{\Delta_{\rm de}}\right)}
\label{eq:wde_2}
\eeq
where $a_{\rm t}$ is the scale factor at which the transition from $w_m$ to $w_0$ takes place and the $\Delta_{\rm de}$ is the steepness of the transition.

\begin{figure}
\centering
\includegraphics[width=0.6\columnwidth]{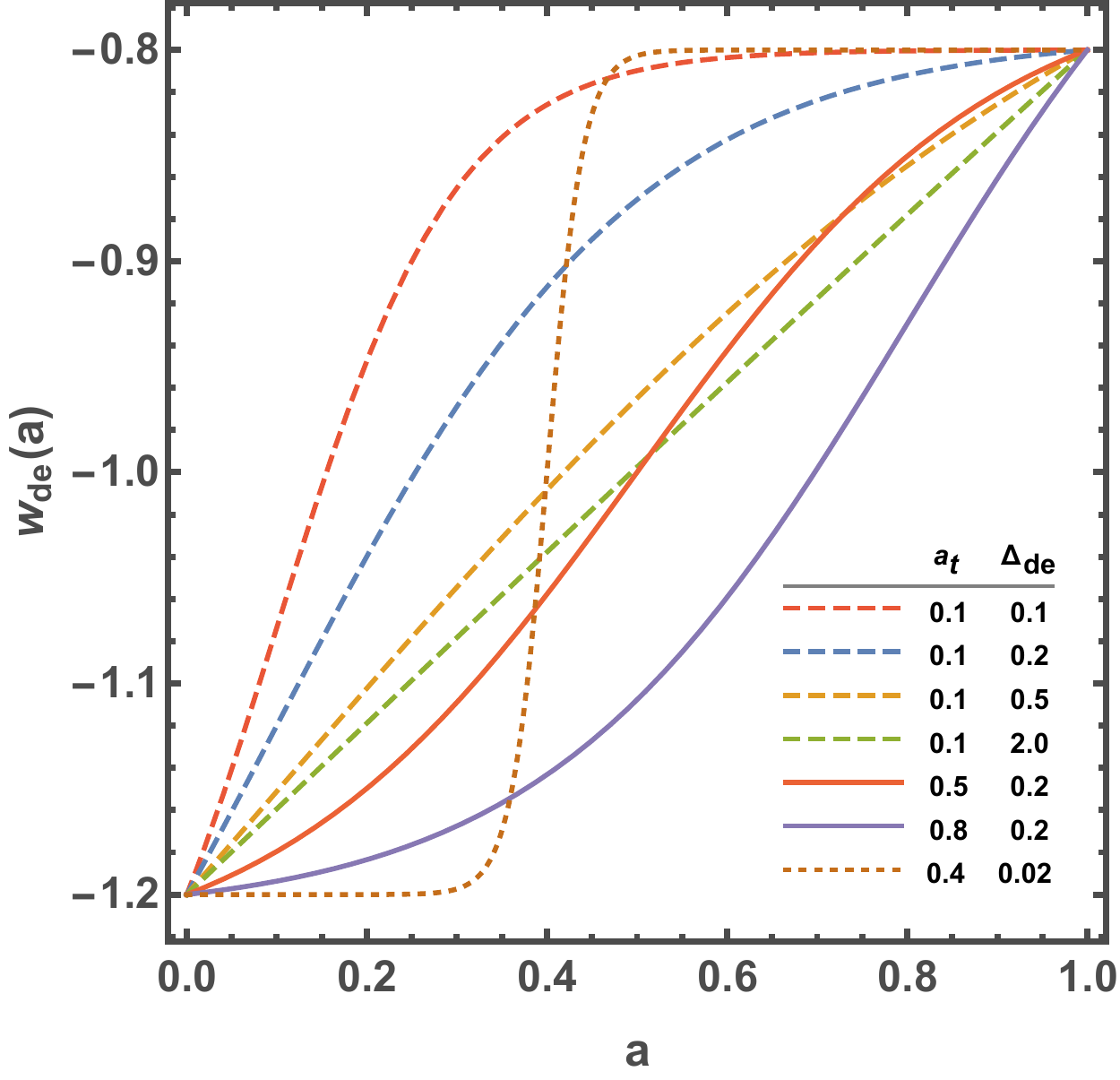}
\caption{Evolution of $w_{\rm de}$ for different sets of values for $a_{\rm t}$ and $\Delta_{\rm de}$, parameters $w_0$ and $w_m$ are fixed to -0.8 and -1.2 respectively.}
\label{fig:wde_evol}
\end{figure}

The factor $\Gamma(a)$ in Eq.~\ref{eq:wde_1}, characterizes the course of the evolution of $w_{\rm de}(a)$. Figure~\ref{fig:wde_evol} elaborates the nature of the parameters $a_{\rm t}$ and $\Delta_{\rm de}$. It can be easily shown that for the two extreme limits of $\Delta_{\rm de}$ Eq.~\ref{eq:wde_1} takes the following form,
\bea
\lim_{\Delta_{\rm de} \to \infty} w_{\rm de}(a) &=& w_0 + (w_m-w_0) \times (1-a) \label{eq:wde_3} \\
\lim_{\Delta_{\rm de} \to 0} w_{\rm de}(a) &=& w_0 + \Theta(a_{\rm t}-a) \times (w_m-w_0) \label{eq:wde_4}
\eea
ie., Equation~\ref{eq:wde_1} approaches to standard 2 parameter parameterization with $w_a = w_m-w_0$ in the limit of $\Delta_{\rm de} \rightarrow \infty$ as this is also evident from the green-dashed plot in the Figure~\ref{fig:wde_evol}. Also when $\Delta_{\rm de} \to 0$ the function $\Gamma(a)$ tend to become a step function (Heaviside function, $\Theta$) around $a=a_{\rm t}$ (the orange-dotted plot in the Figure~\ref{fig:wde_evol}).

Our parameterization is generic to a class of non-interacting scalar field dynamical dark energy models only, ie., we assume $c_{s,{\rm de}}^{\rm eff}=1$. Also this parameterization can only mimic monotonically evolving dynamical dark energy models. 

We will consider a homogeneous and isotropic flat background for the universe described by a FLRW metric. If we neglect the radiation density today, Friedmann equation will have the following form,
\beq
\frac{H^2}{H_0^2} = \Omega_M/a^3 + \Omega_{\rm DE} \exp \left(3 \times \int_{1}^{a} \frac{1+w_{\rm de}(a')}{a'}da' \right)
\label{eq:rho_de}
\eeq
where $\Omega_M$ and $\Omega_{\rm DE}$ is matter density and dark energy density parameters respectively and for a flat universe $\Omega_{\rm DE}+\Omega_{M}=1$.

\subsection{Effect on various cosmological observables}
\subsubsection{Effect on CMB}
\begin{figure}
    \centering
    \includegraphics[scale=0.4]{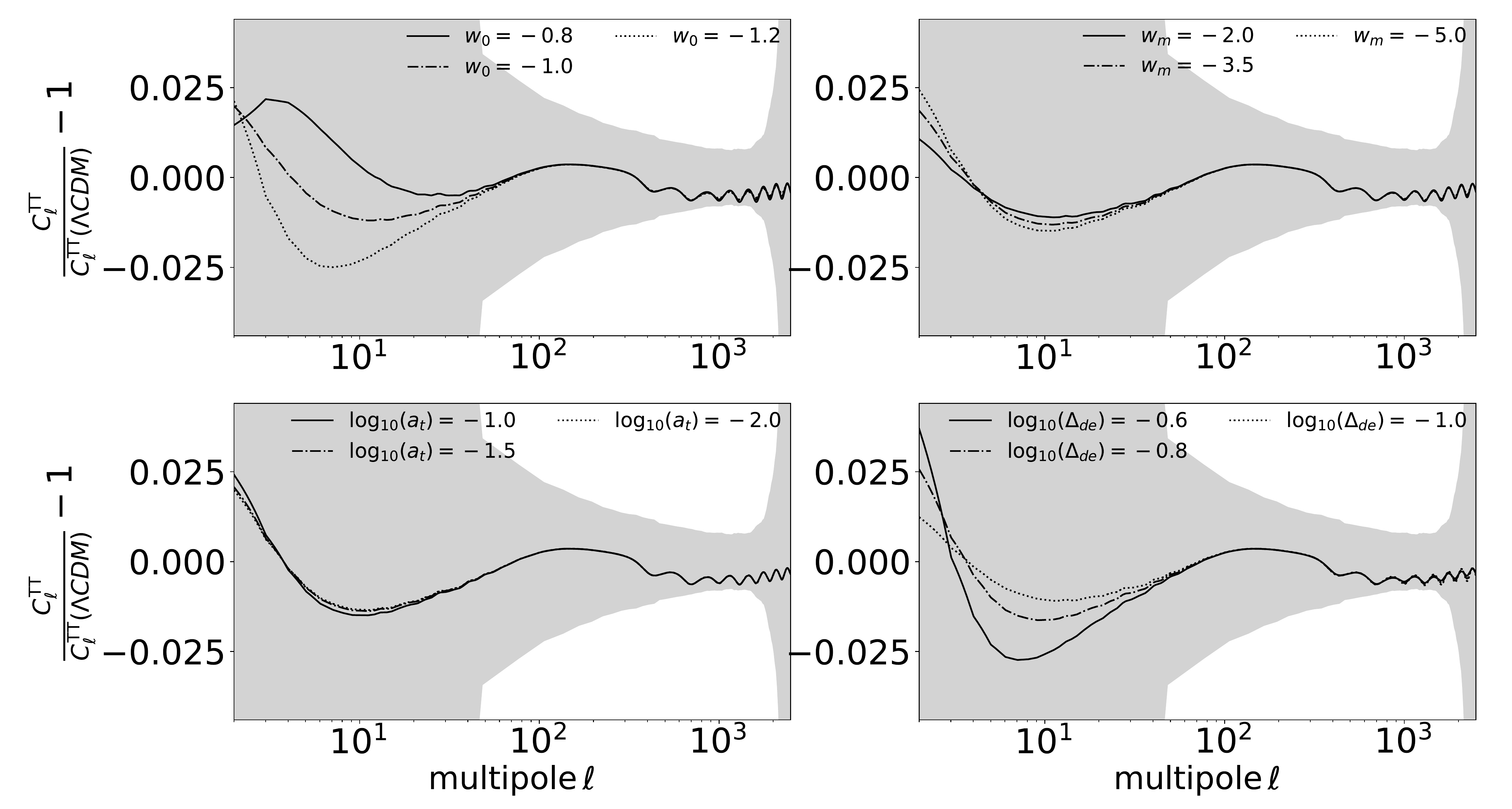}
    \caption{The residual plots of CMB TT power spectra for 4pDE model with respect to $\Lambda$CDM model. The shaded region is the Planck 2018 1$\sigma$ uncertainties. Each of four panels shows effect of varying  different dark energy equation of state parameter (see the legends) while fixing other EoS  parameters to values $(w_0=-1.03,\  w_m=-3.97,\ \log_{10}(a_t)=-1.63,\ \log_{10}(\Delta_{\rm de})=-0.87)$.}
    \label{fig:4pde_cl}
\end{figure}

Though the dark energy energy starts to dominate the energy content of the universe at late times but still it can affect the CMB data in two ways, one is by affecting the angular diameter distance and the second is through Integrated Sachs Wolfe (ISW) effect. The change in the angular diameter distance is reflected in shifts in the location of the peaks in CMB angular power spectra. On the other hand the late time ISW is sensitive to the low $\ell$ regime of the CMB angular power spectra. 
Figure~\ref{fig:4pde_cl} shows the effect of the individual dark energy parameters ($w_0$, $w_m$, $a_t$, $\Delta_{\rm de}$) on the CMB temperature angular power spectrum while keeping the other parameters fixed.
\subsubsection{Effect on BAO}
We know the BAO scale can act as a standard ruler. This scale can be used to measure the angular diameter distance, $d_A(z) = c / (1+z) \ \int_{0}^{z} 1 /H(z') \ {dz'}$, using clustering of galaxies perpendicular to the line of sight and expansion rate $H(z)$ of the universe using clustering along line of sight.
Adjustments in cosmological parameters can alter the clustering scale of the galaxies which is related to the sound horizon $r_d=\int_{z_d}^{\infty} c_s(z)/H(z) \ dz$, where $c_s$ is the sound speed and $z_d$ is the drag epoch. Effectively, BAO measurements actually constrain the combination $r_d*H(z)$ or $d_A(z)/r_d$.
Another important observable in redshift surveys is $f\sigma_8$ which is defined as combination of growth rate $f(a)$ and the root mean square
normalization of the matter power spectrum $\sigma_8$,
$$f\sigma_8(a)=a\frac{\delta'_m(a)}{\delta_m(1)}\sigma_{8,0}$$
Figure~\ref{fig:4pde_hrd} and Figure~\ref{fig:4pde_fsig} in Appdendix respectively show the effect of the individual dark energy parameters ($w_0$, $w_m$, $a_t$, $\Delta_{\rm de}$) on the quantities $r_d \times H(z)/(1+z)$ and $f(z)\times \sigma_8(z)$ while keeping the other parameters fixed.

\subsubsection{Supernovae Ia data}
SNe data provide geometric constraints for dark energy evolution.
These constraints are obtained by comparing the predicted luminosity distance to the SNe with the observed ones. The theoretical model and observations are compared with the measured luminosities, $$M = m - 5 \ {\log_{10}\left(\frac{d_L}{10}\right)}$$
where $M$ is absolute magnitude of the SNe, $m$ is apparent magnitude of the SNe and $d_L= c \ (1+z)\int_{0}^{z} 1 /H(z') \ {dz'}$ is their luminosity distance in parsec. This depends on the evolution of dark energy through $H(z)$. Figure~\ref{fig:4pde_dl} show the effect of the individual dark energy parameters ($w_0$, $w_m$, $a_t$, $\Delta_{\rm de}$) on $d_L(z)$ while keeping the other parameters fixed.

\section{Details of  Analysis}
\subsection{Data Sets}
\label{sec:datasets}
\begin{itemize}
    
    \item Planck 2018 measurements of the low-$\ell$ CMB TT, EE, and  high-$\ell$ TT, TE, EE power spectra, together with the gravitational lensing potential reconstruction \cite{2020A&A...641A...6P}. 
        
    \item The BAO measurements from 6dFGS at $z=0.106$~\cite{Beutler:2011hx}, SDSS DR7 at $z=0.15$~\cite{Ross:2014qpa}, BOSS DR12 at $z=0.38, 0.51$ and $0.61$~\cite{Alam:2016hwk}, and the joint constraints from eBOSS DR14 Ly-$\alpha$ auto-correlation at $z=2.34$~\cite{Agathe:2019vsu} and cross-correlation at $z=2.35$~\cite{Blomqvist:2019rah}.
    
    \item The measurements of the growth function $f\sigma_8(z)$ (FS) from the CMASS and LOWZ galaxy samples of BOSS DR12 at $z = 0.38$, $0.51$, and $0.61$ \cite{Alam:2016hwk}.
    
    \item The Pantheon SNIa catalogue, spanning redshifts $0.01 < z < 2.3$~\cite{Scolnic:2017caz}.
    
    \item  The SH0ES result, modeled with a Gaussian likelihood centered on $H_0 = 73.2 \pm 1.3$ km/s/Mpc \cite{Riess:2020fzl}; however, choosing a different value that {\it combines} various direct measurements would not affect the result, given their small differences.
    
    \item The KIDS1000+BOSS+2dfLenS weak lensing data, compressed as  a split-normal likelihood on the parameter $S_8=0.766^{+0.02}_{-0.014}$ \cite{Heymans:2020gsg}.
     
    \item The Gaussian prior on absolute magnitude $M_B=-19.244 \pm 0.037$ mag \cite{Camarena:2021jlr} , corresponding to the SN measurements from $\rm {SH0ES}$.
     
\end{itemize}

\begin{table}
\centering
  \begin{tabular}{cc}
  \hline
  Parameter & Prior\\
  \hline
  $w_0$&[-5,-0.5]\\
   $w_m$&[-5,-0.5]\\
    $\log(at_{\rm de})$&[-3,0]\\
    $\log(\Delta_{\rm de})$&[-1,0]\\
$\Sigma m_{\nu}$&[0,5]\\
\hline
  \end{tabular}  
  \caption{List of parameter priors for the additional parameters. Priors for the standard base parameters $\{\omega_b,\omega_{\rm cdm},100\times\theta_s,n_s,{\rm ln}(10^{10}A_s),\tau_{\rm reio}\}$ are kept same as the default set into the MontePython-v3 \citep{Brinckmann:2018cvx} code.}
  \label{tab:prior}
\end{table}

\begin{table}
\centering
  \hspace{-2.0cm}
  \begin{tabular}{lcccc}
  \hline
  \multicolumn{1}{c}{ }& \multicolumn{4}{c}{$\nu$4pDE} \\\cmidrule{2-5}
  Data-set $\downarrow$ &  I & II & III & IV \\
  \hline
Planck~high$-\ell$ TT,TE,EE & $2345.70$ &   $2347.49$ & $2347.84$ & $2349.06$ \\
Planck~ low$-\ell$ EE & $396.31$ & $395.78$ & $396.16$ & $396.90$ \\
Planck~ low$-\ell$ TT & $23.56$ & $22.87$ & $22.86$ & $22.76$ \\
Planck~lensing & $8.84$ & $8.95$ & $8.70$ & $9.37$ \\
Pantheon & $1026.90$ & $1027.22$ & $1027.89$ & $1027.93$ \\
BAO~FS~BOSS DR12 & $7.15$ & $7.18$ & $8.52$ & $9.96$ \\
BAO~BOSS low$-z$ & $1.63$ & $2.68$ & $3.245$ & $3.47$ \\
absolute M & $-$ & $13.05$ & $10.43$ & $-$ \\
SHOES & $-$ & $-$ & $-$ & $9.865$ \\
S8 & $-$ & $-$ & $4.717$ & $2.15$ \\
\hline
Total & $3810.13$ & $3825.18$ & $3830.39$ & $3831.50$ \\
\hline
  \end{tabular}  
  \caption{Best-fit $\chi^2$ per experiment (and total) in the $\nu$4pDE model. The column headings in Roman-numeral correspond to different data/prior-combinations as, I) Planck+Ext ; II) Planck+Ext+MB ; III) Planck+Ext+MB+S8; IV) Planck+Ext+H0+S8.}
  \label{tab:chi2_4pde}
\end{table}

\begin{table*}
    \centering
    \hspace{-1.5cm}
    \begin{tabular}{lcccc} 
    \hline
    {} & \multicolumn{4}{c}{$\nu$4pDE} \\
     \cmidrule{2-5}  
     Parameter & Planck+Ext & Planck+Ext+$M_B$ & Planck+Ext+$M_B$+$S_8$ &Planck+Ext+$H_0$+$S_8$ \\
     \hline 
     $100~\omega_{b}$ &  $2.231(2.235)_{-0.01359}^{+0.01423}$  & $2.235(2.247)_{-0.01479}^{+0.01554}$& $2.239(2.249)_{-0.0148}^{0.0155}$
    & $2.238(2.239)_{-0.015}^{+0.015}$\\
     $\omega_{\rm cdm}$ &  $0.1202(0.1202)_{-0.0010613}^{+0.00108}$  & $0.1200(0.1185)_{-0.0010872}^{+0.0010481}$& $0.1193(0.1191)_{-0.00109}^{0.00105}$
     & $0.1194(0.1192)_{-0.001}^{+0.0011}$ \\
     $100*\theta_{s }$ & $1.0418(1.0417)_{-0.0003067}^{+0.00028}$  & $1.0419(1.0419)_{-0.0002733}^{+0.0003093}$& $1.0419(1.0419)_{-0.00273}^{0.000309}$
     & $1.042(1.04198)_{-0.00028}^{+0.00029}$ \\
     $n_{s }$ &$0.9631(0.9642)_{-0.004134}^{0.0038948}$   & $0.9636(0.9679)_{-0.0042631}^{0.004.000}$& $0.9647(0.9652)_{-0.00426}^{0.004}$& $0.9645(0.9644)_{-0.0042}^{+0.0042}$ \\
     ${\rm ln}(10^{10}A_{s })$ &  $3.043(3.047)_{-0.0147}^{+0.0150}$  & $3.042(3.037)_{-0.014812}^{+0.01488}$& $3.042(3.040)_{-0.0148}^{0.0149}$  & $3.042(3.0469)_{-0.016}^{+0.015}$ \\
     $\tau_{\rm reio }$ & $0.0537(0.0555)_{-0.007351}^{+0.007626}$   & $0.05357(0.05308)_{-0.007737}^{+0.0072268}$& $0.05436(0.05511)_{-0.00754}^{+0.00723}$ 
     & $0.054(0.0577)_{-0.0079}^{+0.0076}$ \\
     $\Sigma m_{\nu}$ [eV] &  $0.1129(0.038)_{-0.1129}^{0.0276}$  & $0.1069(0.0094)_{-0.1069}^{0.0697}$& $0.1847(0.1043)_{-0.165}^{0.0698}$ & $0.1857(0.2748)_{-0.13}^{+0.091}$ \\

     $w_0$ &  $-0.9856(-1.038)_{-0.065}^{+0.053}$  & $-0.9886(-1.011)_{-0.073}^{+0.053}$& $-0.9901(-1.029)_{-0.0766}^{0.0561}$ &$-1.04(-0.9446)_{-0.062}^{+0.049}$ \\
     $w_m$ &  $-2.285(-1.040)_{-0.59}^{+1.5}$  & $-2.715(-1.611)_{-0.91}^{+1.6}$& unconstrained&$-2.45(-3.1)_{-0.78}^{+1.4}$ \\
     $\log_{10}(a_{\rm t})$ &  unconstrained  & unconstrained & unconstrained & unconstrained \\
     $\log_{10}(\Delta_{\rm de})$ &  $-0.6752(-0.7649)_{-0.32}^{+0.066}$  & $-0.6687(-0.9264)_{-0.33}^{+0.065}$& $-0.6313(-0.8739)_{-0.368}^{0.0809}$  & $-0.6936(-0.4452)_{-0.31}^{+0.068}$ \\
    \hline 
    $M_{B}$ &  $-19.40(-19.40)_{-0.01962}^{+0.01887}$& $-19.37(-19.37)_{-0.01655}^{+0.01629}$&   $-19.369(-19.359)_{-0.01615}^{+0.01580}$ & $-19.37(-19.36)_{-0.015}^{+0.017}$ \\
    $\sigma_{8}$ &  $0.8135(0.8279)_{-0.01203}^{+0.01685}$  & $0.8225(0.8216)_{-0.01303}^{+0.01658}$& $0.8073(0.8206)_{-0.01462}^{+0.01763}$ & $0.8093(0.7966)_{-0.014}^{+0.017}$ \\
     $\Omega_{m }$ &  $0.3064(0.3038)_{-0.00810}^{+0.00779}$  & $0.2973(0.2954)_{-0.0074}^{+0.0069}$&$0.294(0.2893)_{-0.00654}^{0.00637}$& $0.2919(0.2955)_{-0.0064}^{+0.006}$ \\
    $S_{8 }$ &  $0.822(0.832)_{-0.0117}^{+0.0170}$  & $0.8188(0.81)_{-0.013}^{+0.016}$& $0.799(0.8051)_{-0.0134}^{0.0163}$ & $0.7982(0.8043)_{-0.013}^{+0.014}$ \\
    $H_0$ [km/s/Mpc] & $68.21(68.53)_{-0.823}^{+0.846}$ & $69.22(69.14)_{-0.791}^{+0.843}$& $69.44(69.87)_{-0.724}^{0.697}$ & $69.71(69.58)_{-0.7}^{+0.7}$ \\
    \hline 
    $\chi^2_{\rm min}$ & 3810.13 & 3825.18   &3830.34  &3831.50 \\
    \hline 
    $\Delta \chi^2_{\rm min}$ & -1.4 & -6.53   & -5.5&-4.3 \\
    \hline
    \end{tabular}
    \caption{The mean (best-fit) $\pm1\sigma$ error of the cosmological parameters reconstructed from the lensing-marginalized Planck+BAO+SN1a data and combinations of $M_B$ and $S_8$ priors for $\nu$4pDE model. We also report the corresponding $\Delta \chi^2_{\rm min}$ values.}
    \label{tab:MCMC2}
\end{table*}

\begin{table*}
    \centering
    \begin{tabular}{lccc}
    \hline
    { } & \multicolumn{3}{c}{$\nu$CPL} \\
    \cmidrule{2-4} 
    Parameter & Planck+Ext & Planck+Ext+$M_B$ & Planck+Ext+$S_8$+$M_B$ \\
     \hline         
     $100~\omega_{b}$ &  $2.2334(2.23425)_{-0.01418}^{+0.01460}$  & $2.2364(2.2461)_{-0.01442}^{+0.01429} $&$2.239(2.2291)_{-0.015}^{+0.015}$  \\
     $\omega_{\rm cdm}$ &  $0.12016(0.1197)_{-0.001073}^{+0.001131}$   & $0.1201(0.1196)_{-0.0011}^{+0.0011}$&
    $0.1194(0.1189)_{-0.001}^{+0.0011}$  \\
     $100*\theta_{s }$ & $1.0419(1.197716)_{-0.0003023}^{+0.0002927}$   & $1.042(1.04173)_{-0.00028}^{+0.0003}$&$1.042(1.0418)_{-0.00031}^{+0.00028}$ \\
     ${\rm ln}(10^{10}A_{s })$ &  $3.0448(1.197716)_{-0.01570}^{+0.01478}$ & $3.043(3.0378)_{-0.015}^{+0.015}$& $3.041(3.0268)_{-0.015}^{+0.015}$\\
     $n_{s }$ &$0.96397(0.9657)_{-0.00394}^{+0.00416} $ & $0.9641(0.9657)_{-0.0041}^{+0.0041}$& $0.9645(0.9669)_{-0.0042}^{+0.0041}$\\
     $\tau_{\rm reio }$ & $0.05446(0.05415)_{-0.00801}^{+0.00752}$ & $0.05375(0.05149)_{-0.0077}^{+0.0074}$&  $0.05388(0.04686)_{-0.0074}^{+0.0073}$ \\
     $\Sigma m_\nu$ [eV] &  $0.1092(0.04848)_{0.100}^{+0.0255}$ & $0.09942(0.00578)_{-0.099}^{+0.025}$& $0.1877(0.074)_{-0.14}^{+0.094}$\\

     $w_0$ &  $-0.9690(-0.99195)_{-0.0830}^{+0.0787}$ & $-0.9541(-0.9558)_{-0.094}^{+0.076}$&  $-0.9393(-1.08)_{-0.095}^{+0.09}$  \\
     $w_a$ &  $-0.2915(-0.1680)_{-0.292}^{+0.409}$ & $-0.5008(-0.3023)_{-0.28}^{+0.49}$& $-0.6716(0.0303)_{-0.39}^{+0.55}$ \\
    
    \hline 
    $M_B$ &  $-19.40(-19.3923)_{-0.0179}^{+0.0191}$ & $-19.37(-19.374)_{-0.016}^{+0.016}$&   $-19.37(-19.376)_{-0.017}^{+0.016}$ \\
    $\sigma_{8}$ &  $0.8134(0.8237)_{-0.0120}^{+0.0155}$ & $0.8237(-0.8306)_{-0.012}^{+0.016}$&  $0.8061(0.8172)_{-0.015}^{+0.018}$\\
     $\Omega_{m }$ &  $0.3074(0.3017)_{-0.00808}^{+0.00723}$ & $0.2974(0.2977)_{-0.00718}^{+0.00664}$&$0.2948(0.2860)_{-0.00703}^{+0.00673}$ \\
    $S_{8}$ &  $0.8233(0.827)_{-0.0124}^{+0.0156}$  & $0.820(0.825)_{-0.0125}^{+0.0152}$& $0.798(0.802)_{-0.0127}^{+0.0157}$ \\
    $H_0$ [km/s/Mpc] & $68.09(68.62)_{-0.768}^{+0.824}$  & $69.23(69.10)_{-0.742}^{+0.740}$&  $69.36(69.71)_{-0.772}^{+0.720}$ \\
    \hline 
    $\chi^2_{\rm min}$ &3809.86  & 3824.43   & 3833.08  \\
    \hline 
    $\Delta \chi^2_{\rm min}$ & -1.67 & -7.23   & -2.76  \\
    \hline 
    \end{tabular}
    \caption{The mean (best-fit) $\pm1\sigma$ error of the cosmological parameters reconstructed from the lensing-marginalized Planck+BAO+SN1a data and combinations of $M_B$ and $S_8$ priors for $\nu$CPL model. We also report the corresponding $\Delta \chi^2_{\rm min}$ values.}
    \label{tab:MCMC3}
\end{table*}

\begin{figure*}
\centering
\includegraphics[width=0.7\columnwidth]{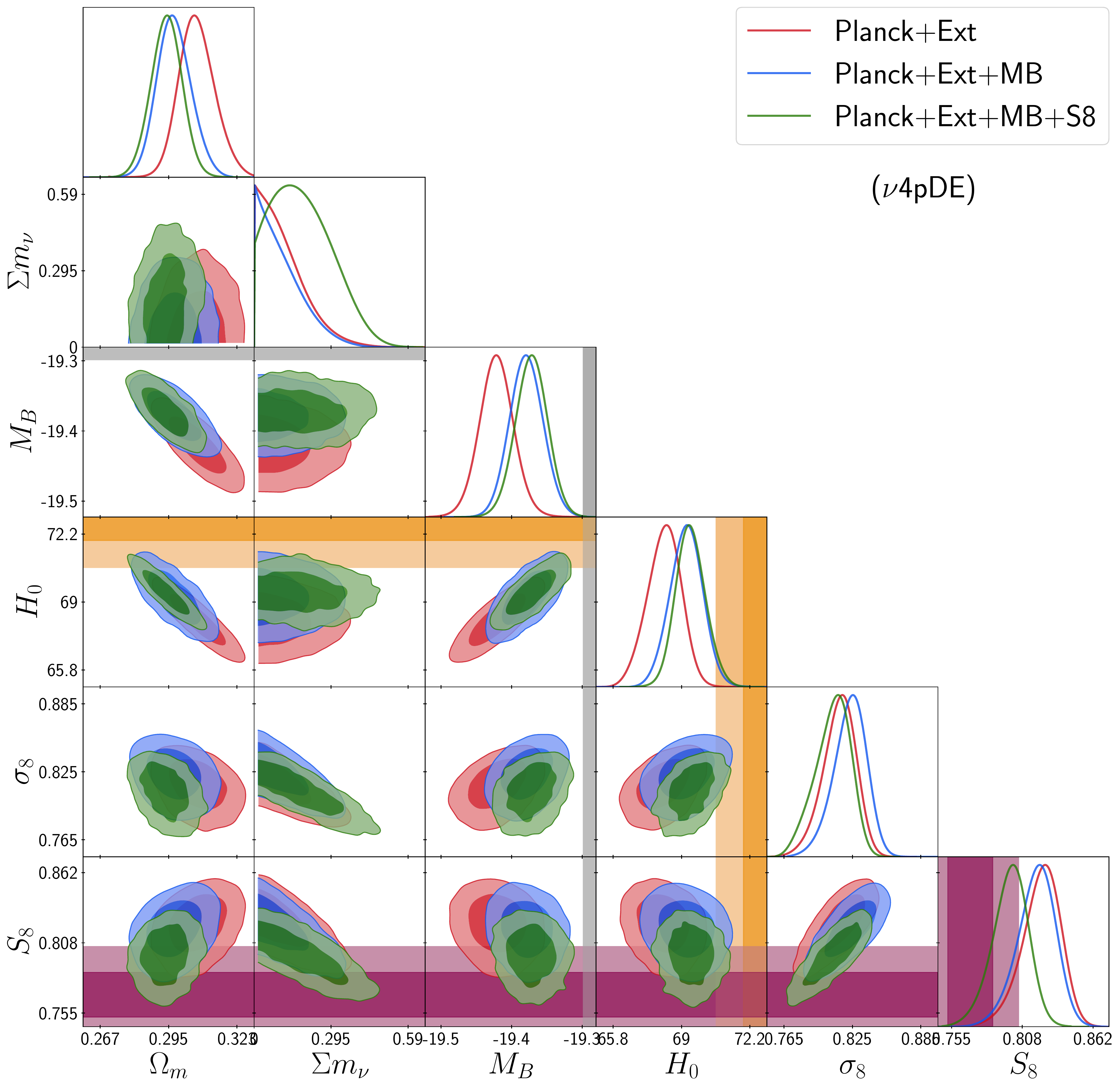}
\caption{2D posterior distributions of ($\Omega_{\rm m}$, $\Sigma m_{\nu}$, $M_B$, $H_0$, $\sigma_8$, $S_8$) for $\nu$4pDE model with Planck+Ext data and different prior combination. We have also added 68\% (dark orange) and 95\% (light orange) bands corresponding to a $H_0$ measurement from SH0ES and  68\% (dark purple) and 95\% (light purple) bands corresponding to $S_8$ measurement from KIDS1000+BOSS+2dfLenS.}
\label{fig:4pde}
\end{figure*}

\begin{figure*}
\centering
\includegraphics[width=0.7\columnwidth]{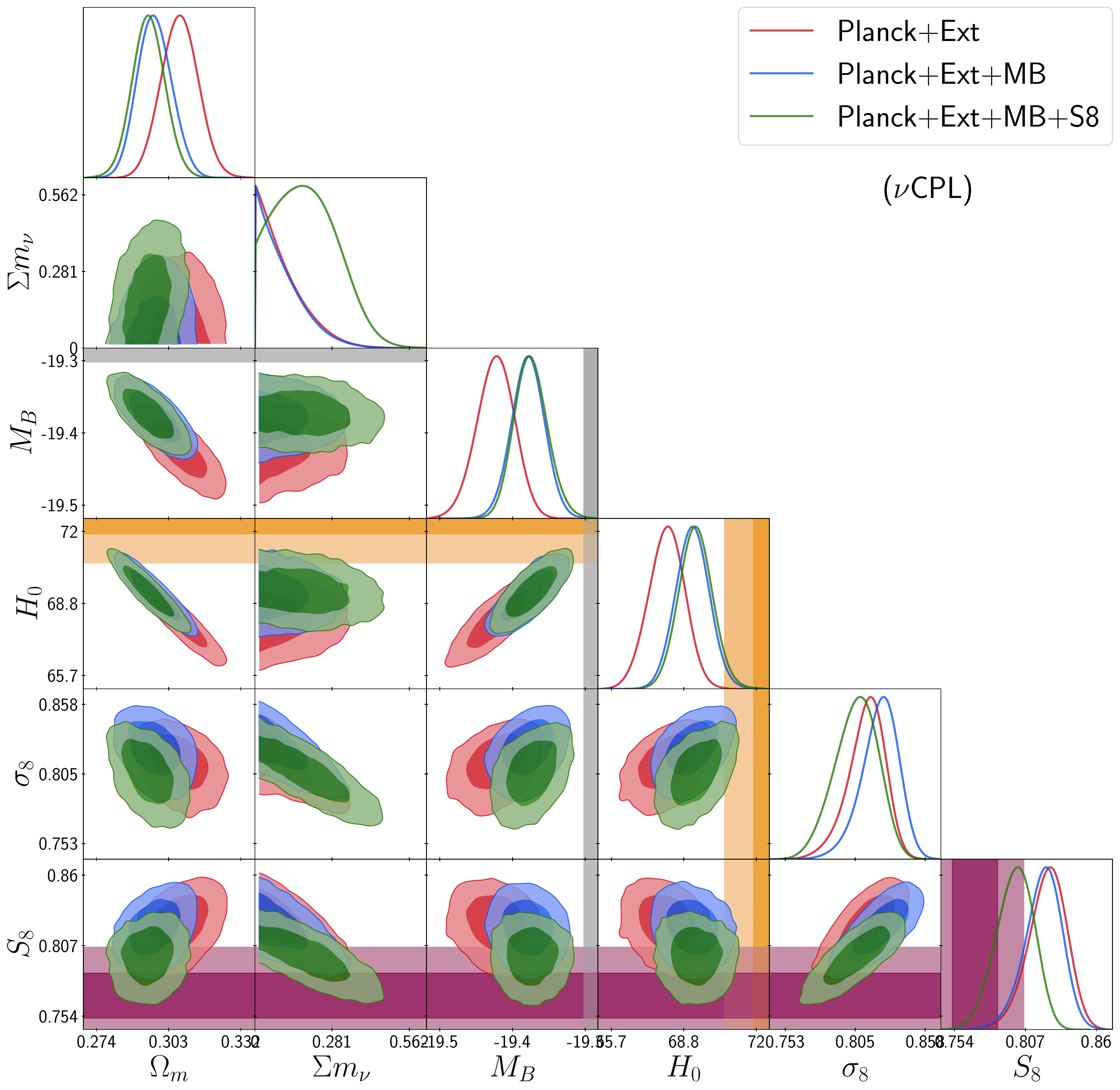}

\caption{2D posterior distributions of ($\Omega_{\rm m}$, $\Sigma m_{\nu}$, $M_B$, $H_0$, $\sigma_8$, $S_8$) for $\nu$CPL model with Planck+Ext data and different prior combination. We have also added 68\% (dark orange) and 95\% (light orange) bands corresponding to a $H_0$ measurement from SH0ES and  68\% (dark purple) and 95\% (light purple) bands corresponding to $S_8$ measurement from KIDS1000+BOSS+2dfLenS.}
\label{fig:clp}
\end{figure*}

\begin{figure*}
\centering
\includegraphics[width=0.65\columnwidth]{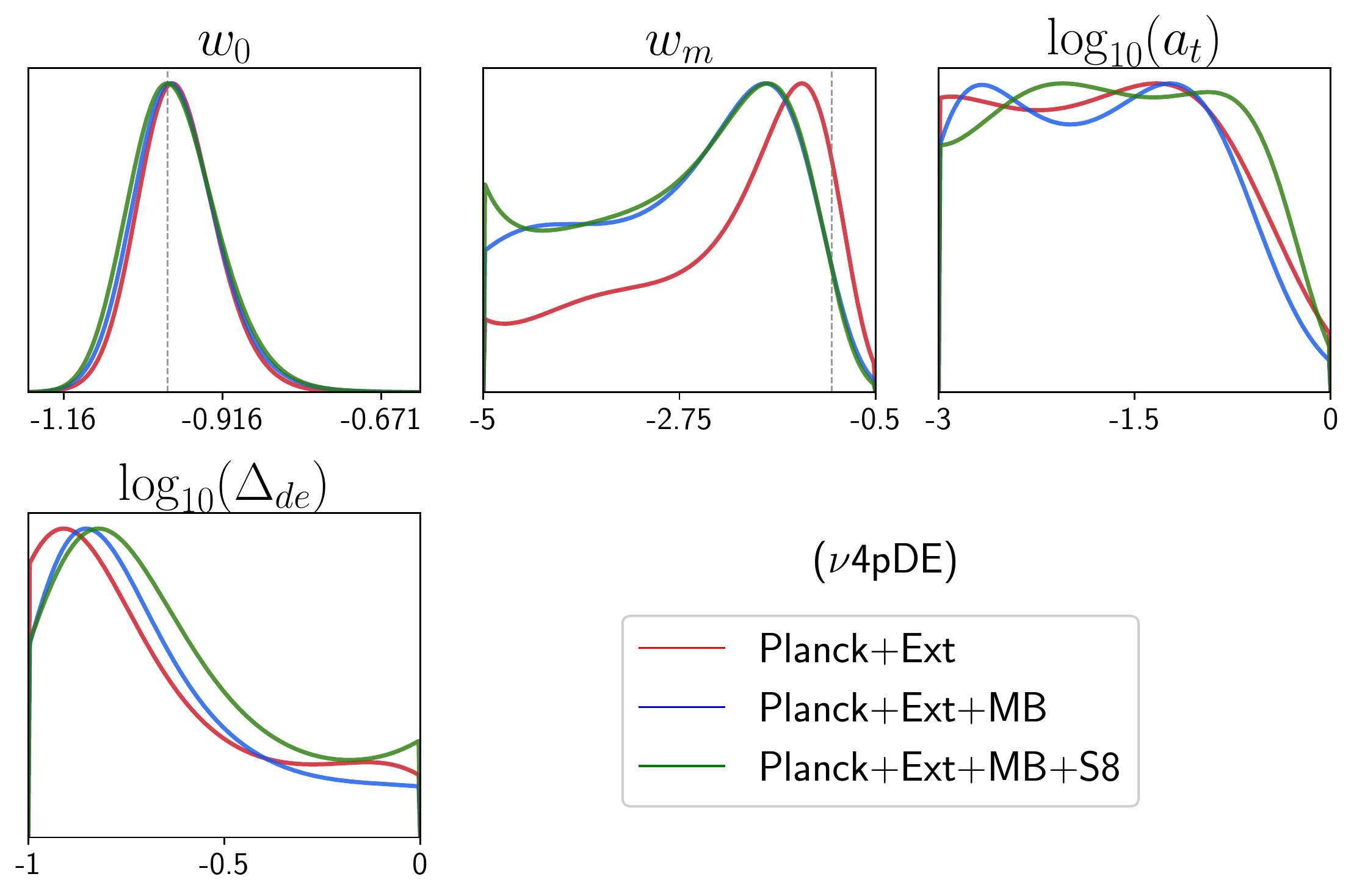}

\caption{1D posterior distributions of equation of state parameters \{$w_0$, $w_m$,  $\log_{10}(a_{\rm t})$, $\log_{10}(\Delta_{\rm de})$\} for $\nu$4pde model with Planck+Ext data and different prior combinations (see legends).}
\label{fig:eos_bao}
\end{figure*}

\begin{figure}
\centering
\includegraphics[width=0.64\columnwidth]{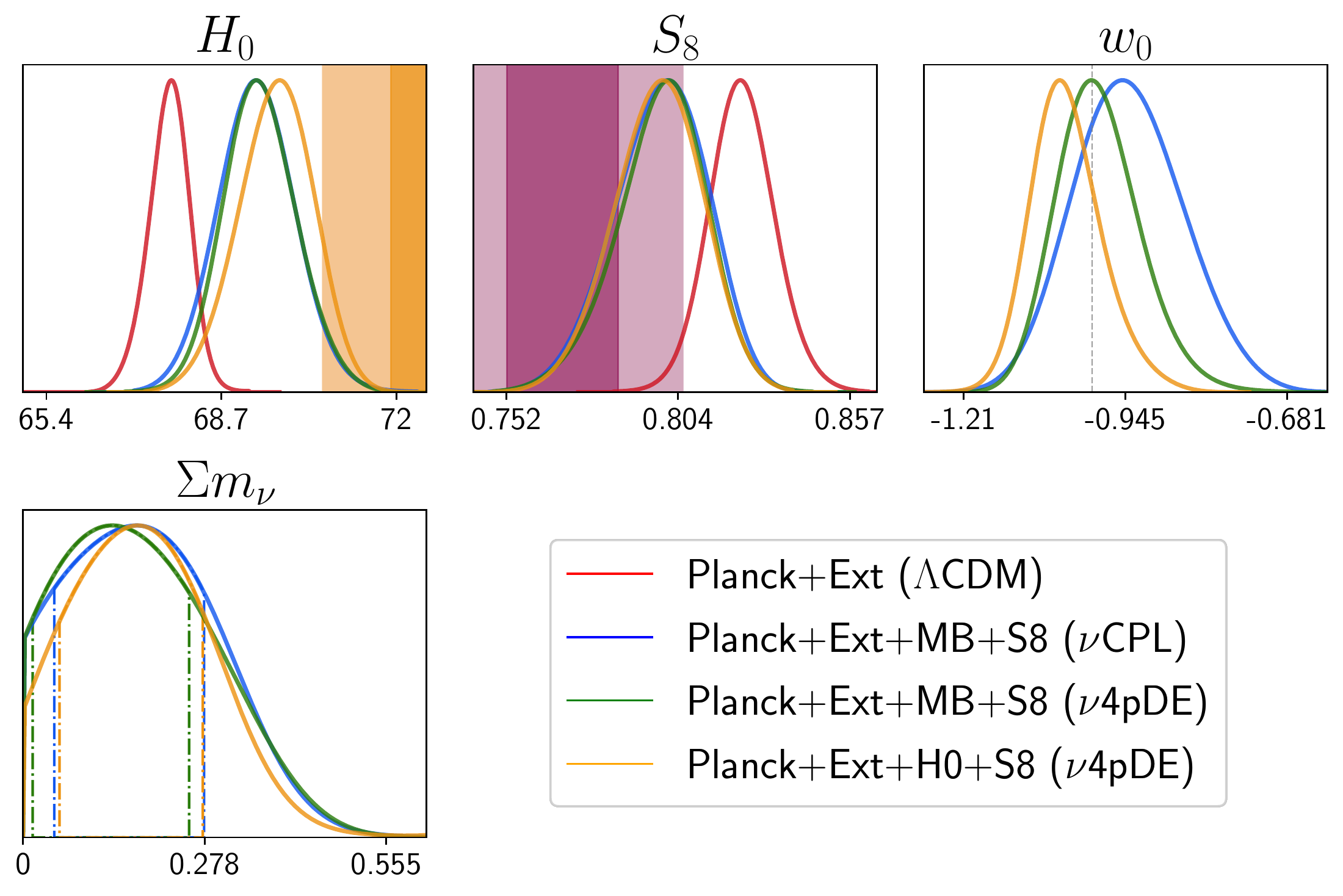}
\caption{1D posterior distributions of ($H_0$, $S_8$, $w_0$, $\sum m_{\nu}$) for different models and combination of priors (see legends). The dot-dashed vertical lines in the $\Sigma m_\nu$ panel correspond to respective 1 $\sigma$ levels.}
\label{fig:compare_1d}
\end{figure}

\subsection{Methodology}
Our baseline cosmology consists in the following combination of the six $\Lambda$CDM parameters $\{\omega_b,\omega_{\rm cdm},100\times\theta_s,n_s,{\rm ln}(10^{10}A_s),\tau_{\rm reio}\}$, plus four dark energy equation of state parameters as discussed in Section~\ref{sec:model}, namely {$\rm w_0$, $\rm w_m$, $a_{\rm t}$, $\Delta_{\rm de}$} and neutrino mass $\Sigma m_{\nu}$. We dub this model as $\nu$4pDE. We run MCMC analysis of $\nu$4pDE model against various combinations of the CMB, BAO and supernovae data sets (details of which is given in Section~\ref{sec:datasets}) with the Metropolis-Hasting algorithm as implemented in the MontePython-v3 \citep{Brinckmann:2018cvx} code interfaced with our modified version of CLASS. We use the prior ranges for $\nu$4pDE model as given in the Table~\ref{tab:prior}, priors for the other six base parameters $\{\omega_b,\omega_{\rm cdm},100\times\theta_s,n_s,{\rm ln}(10^{10}A_s),\tau_{\rm reio}\}$ are kept same as the default set into the MontePython-v3 \citep{Brinckmann:2018cvx} code. 
All reported $\chi^2_{\rm min}$ are obtained with the python package {\sc iMinuit \footnote{\url{https://iminuit.readthedocs.io/}}} \cite{James:1975dr}. We make use of a Choleski decomposition to better handle the large number of nuisance parameters \cite{Lewis:1999bs} and consider chains to be converged with the Gelman-Rubin convergence criterion $R-1<0.05$ \cite{Gelman:1992zz}. We also run MCMC chains for the standard $\Lambda$CDM and $\nu$CPL (CPL parameterization with neutrino mass as free parameter) models with different combinations of data sets and priors for making comparisons.

\section{Results}
\label{sec:results}
We ran three sets of models, the first one is the standard ``$\Lambda$CDM Model". The other two are dynamical dark energy models namely $\nu$4pDE and $\nu$CPL models model. 
Each model is constrained with combinations of the data sets, Planck TT, EE, TE+Planck Lensing (see Section~\ref{sec:datasets}) dubbed as ``Planck", BAO and Pantheon (see Section~\ref{sec:datasets}), together dubbed as ``Ext" and the priors $H_0$, $M_B$ and $S_8$ (see Section~\ref{sec:datasets}) dubbed as ``H0'', ``MB'' and ``S8'' respectively. We use the standard $\Lambda$CDM model with Planck+Ext dataset as our base model for computing $\Delta \chi^2$ values.

The results of the $\nu$4pDE model with combined data sets for various cases are reported in Table~\ref{tab:MCMC2}, the 2D posterior distributions ($H_0$, $M_B$, $S_8$ and $\Sigma m_\nu$) are shown in Figure~\ref{fig:4pde} and the 1d posterior distributions of the dark energy equation (DE) of state parameters are shown in the Figure~\ref{fig:eos_bao}. The $\chi^2_{\rm min}$ values for this model corresponding to different dataset/prior combinations are reported in Table~\ref{tab:chi2_4pde}.

We find that $w_0$ is well constrained for each dataset combination and is consistent with the cosmological constant. However, the other three DE parameters are less constrained or unconstrained. 
Especially in the case of parameter $\log_{10}(\rm a_{\rm t})$ we don't find the lower bound.  
This is not surprising given the fact the data does not seem to be favoring very sharp transition, ie., very small value for $\Delta_{\rm de}$, in that case $w_{\rm de}(z)$ has rather weak dependence on the parameter $a_t$ (see Equations~\ref{eq:wde_3} and \ref{eq:wde_4}) and hence resulting in a weak constraints on the parameter $a_t$.
The overall $\Delta \chi^2_{\rm min}$ is -1.7 compared to our base $\Lambda$CDM model, though  we still have the $H_0$ tension at $\sim 3.2\sigma$ and $S_8$ tension at $\sim 2.5\sigma$ with this model.

When using the $M_B$ prior, there is no major impact on the equation of state parameters except the values of $w_m$ shift slightly more negative. In this case, the model has $H_0$ tension at $\sim 2.6\sigma$  with SH0ES results. The overall $\chi^2_{\rm min}$ shift is -6.5 compared to the $\Lambda$CDM model.

For the case when $M_B$ and $S_8$ priors are applied simultaneously, we find that $H_0$  attains high value, 
and there is a decrease in $S_8$ value ie., we find negative correlation between parameter $S_8$ and $H_0$ in this case (see Figure~\ref{fig:4pde}). This model brings down the $S_8$ tension below $\leq 1.5\sigma$ and $H_0$ tension from $\leq 2.5 \sigma$. In this case, we also get a peak in the posterior of the neutrino mass (see Figure~\ref{fig:4pde}). 

\subsection{Comparison between  \texorpdfstring{$\nu$}{}CPL Model and the  \texorpdfstring{$\nu$}{}4pDE model} 
We also run the $\nu$CPL model with same data combinations. Results of this model reported in Table~\ref{tab:MCMC3} and the 2D posterior distributions are shown in Figure~\ref{fig:clp}. 
Both the parameters $w_0$ and $w_a$ are well constrained for $\nu$CPL model. We find the posterior distribution of main cosmological parameters \{$\omega_b,\omega_{\rm cdm}, 100\times\theta_s, n_s, {\rm ln}(10^{10}A_s), \tau_{\rm reio}$\} of this model are matched with the $\nu$4pDE model. However the model parameters ($w_0$ and $w_a$) are different for obvious reasons.

Figure~\ref{fig:compare_1d} shows a comparison among the $\nu$4pDE, $\nu$CPL and the base $\Lambda$CDM model in terms of the 1d posteriors of parameters \{$H_0$, $S_8$, $w_0$, $\sum m_{\nu}$\}. The parameter $w_0$ is constrained more in the $\nu$4pDE model compare to the $\nu$CPL model. However, when we don't use any prior, we do not notice a significant change in $H_0$ and $S_8$ compared to $\Lambda$CDM model. But when we use $M_B$ prior, the level of $H_0$ tension is reduced, and is within  2.5$\sigma$ level with SH0ES measurement. Similarly, when we add $S_8$ prior, we notice a slight reduction in the $S_8$ parameter also. 

\subsection{Comparison between  \texorpdfstring{$H_0$}{} prior and the  \texorpdfstring{$M_B$}{} prior}
When using the prior on $H_0$ instead of $M_B$, the main impact on results is on the parameters $w_0$ and $H_0$.
The parameter $H_0$ attains a slightly higher value compared to $M_B$ prior case and 
$w_0$ shifts towards a lower value. The results are compared in Figure~\ref{fig:compare_1d}. This result is true for $\nu$CPL model as well. We also see that $H_0$ prior supports a non zero neutrino mass more in comparison to the $M_B$ prior.
In summary we say that use of $H_0$ prior has stronger impact on the $w_0$, $\Sigma m_\nu$ and for obvious reasons on $H_0$ compared to the case with $M_B$ prior.

\section{Conclusions and discussion}
\label{sec:discussion}

We have investigated a generic model of non-linearly evolving dynamical dark energy
equation of state $w$ with massive neutrinos ($\nu$4pDE model), in the light of the recent Planck, Pantheon, BAO data and recent measurements of $H_0$ and $S_8$. We have studied the effect of such parameterization on the background and derived cosmological parameters. The results of our analysis is in accordance with the earlier studies with a linear evolution of $w_{\rm de}$ (CPL like) \cite{2018PhRvD..97l3504P,2017PhRvD..96b3523D,2016PhLB..761..242D,2015PhRvD..92l1302D} and a few other parameterization of dynamical dark energy \cite{2017NatAs...1..627Z,2017RAA....17...50Z,2019ApJ...876..143B}. We do not get any strong constraints on the two extra parameters $a_{\rm t}$ and $\Delta_{\rm de}$ at $\gtrsim 1\sigma$ level, constraints on $w_0$  are in good agreement with the $\nu$CPL parameterization. However, we do get slightly different and a more tightly constrained $w_0$ when the model is tested against Planck and external data (Pantheon+BAO) put together. The added external prior on parameters $M_B$ and $S_8$ makes the difference in $\nu$CPL and $\nu$4pDE parameterization even broader. The $\nu$4pDE model can bring down the Hubble tension to $\sim 2.5 \sigma$ level and the $S_8$ tension to $\sim 1.5 \sigma$ level when tested against Planck, BAO and Pantheon supernovae data together. More importantly, we find that there is a slightly negative correlation between the parameters $S_8$ and $H_0$ with $\nu$4pDE model which is very interesting.  

However, both the $\nu$4pDE model and $\nu$CPL model improves $\Delta \chi^2_{\rm min}$  for the Planck+Ext dataset and the recent measurements of $H_0$ and $S_8$  in comparison to $\Lambda$CDM, this lowering of $\chi^2_{\rm min}$ is achieved at the expense of adding extra  parameters. So if we follow $\Delta$AIC criteria\footnote{Akaike Information Criterion (AIC) is one of the popular methods of estimating the relative quality of proposed models for a given data. AIC is based on using a trade-off between the goodness of fit of the model and the simplicity. AIC uses a model’s log-likelihood as a measure of fit and the number of parameters in the model as the complexity of the model. If $N_{\rm Model}$ is the total number of parameters in a model the AIC score for that model is given by,
$${\Delta \rm AIC} = {\chi^2_{\rm min,Model}} - {\chi^2_{\rm min,\Lambda CDM}}+2(N_{\rm Model}-N_{\Lambda \rm CDM})$$}, the level of success of these models degrades as none of the models has significantly improved $\Delta$AIC value over $\Lambda$CDM model. 
From table \ref{tab:AIC} it is evident that when considered Planck+Ext+MB, $\nu$CPL is still a preferred model. But when we consider Planck+Ext+MB+S8, all the models (4pDE, $\nu$4pDE, $\nu$CPL) perform worse in comparison to $\Lambda$CDM.

\begin{table}
\centering
  \begin{tabular}{l|ccccc}
  \hline
 
  \multicolumn{1}{l|}{ Data $\rightarrow$}& \multicolumn{2}{c|}{Planck+Ext+MB}& \multicolumn{2}{c}{Planck+Ext+MB+S8} \\
  \hline
  Model $\downarrow$ &\multicolumn{1}{c}{$\Delta \chi^2_{\rm min}$} &\multicolumn{1}{c|} {$\Delta \rm{AIC}$} &\multicolumn{1}{c}{$\Delta \chi^2_{\rm min}$} &\multicolumn{1}{c}{$\Delta \rm{AIC}$}\\
  \hline
  $\Lambda$CDM&0&0&0&0\\
  $\nu$CPL&$-7.23$&$-1.23$&$-2.76$&$+3.24$\\
  $\nu$4pDE&$-6.53$&$+3.47$&$-5.50$&$+4.50$\\
  \hline
  \end{tabular}  
  \caption{Comparison of $\Delta \chi^2_{\rm min}$ and $\Delta \rm {AIC}$ for $\nu$4pDE and $\nu$CPL models.}
  \label{tab:AIC}
\end{table}



We also see that with added $S_8$ prior $\nu$4pDE favours a non-zero value for the neutrino mass parameter ($\Sigma m_\nu \sim 0.2 \pm 0.1$ eV), which is in agreement  with earlier work using the $\nu$CPL parameterization \citep{Poulin:2018zxs} however our analysis does not detect lower bound on $\Sigma m_\nu$ parameter at $\gtrsim 1 \sigma$. Earlier work such as  \cite{Poulin:2018zxs} where 2015 Planck data has been used, the neutrino mass $\Sigma m_\nu$ was found to be  non-zero even at $\gtrsim 2 \sigma$. Also in a recent work \cite{2021arXiv211202993D} reports a non-zero value for the neutrino mass when used WMAP data along with  ACT-DR4 and SPT-3G data. However they also get similar results on $\Sigma m_\nu$ as ours when recent Planck and BAO data is used. The well known lensing anomaly {in the Planck data} ($A_{\rm lens} \ne 1$) could be playing a role here which needs to be investigated further in more details.

\section*{Acknowledgements}

 We thank  Vivian Poulin for reading the draft and for his valuable inputs.  We thank  Pier Stefano Corasaniti , Eoin O Colgain and Sunny Vagnozzi for their valuable inputs.  SD and KP acknowledge SERB DST Government of India grant  CRG/2019/006147 for supporting the project. We acknowledge HPC NOVA, IIA Bangalore where numerical simulations were performed.



\clearpage
\bibliographystyle{aasjournal}
\bibliography{main} 




\appendix

\section{Effect of varying 4pDE Equation of state parameters on BAO and supernovae observables}

\begin{figure}[hbt!]
    \centering
    \includegraphics[scale=0.38]{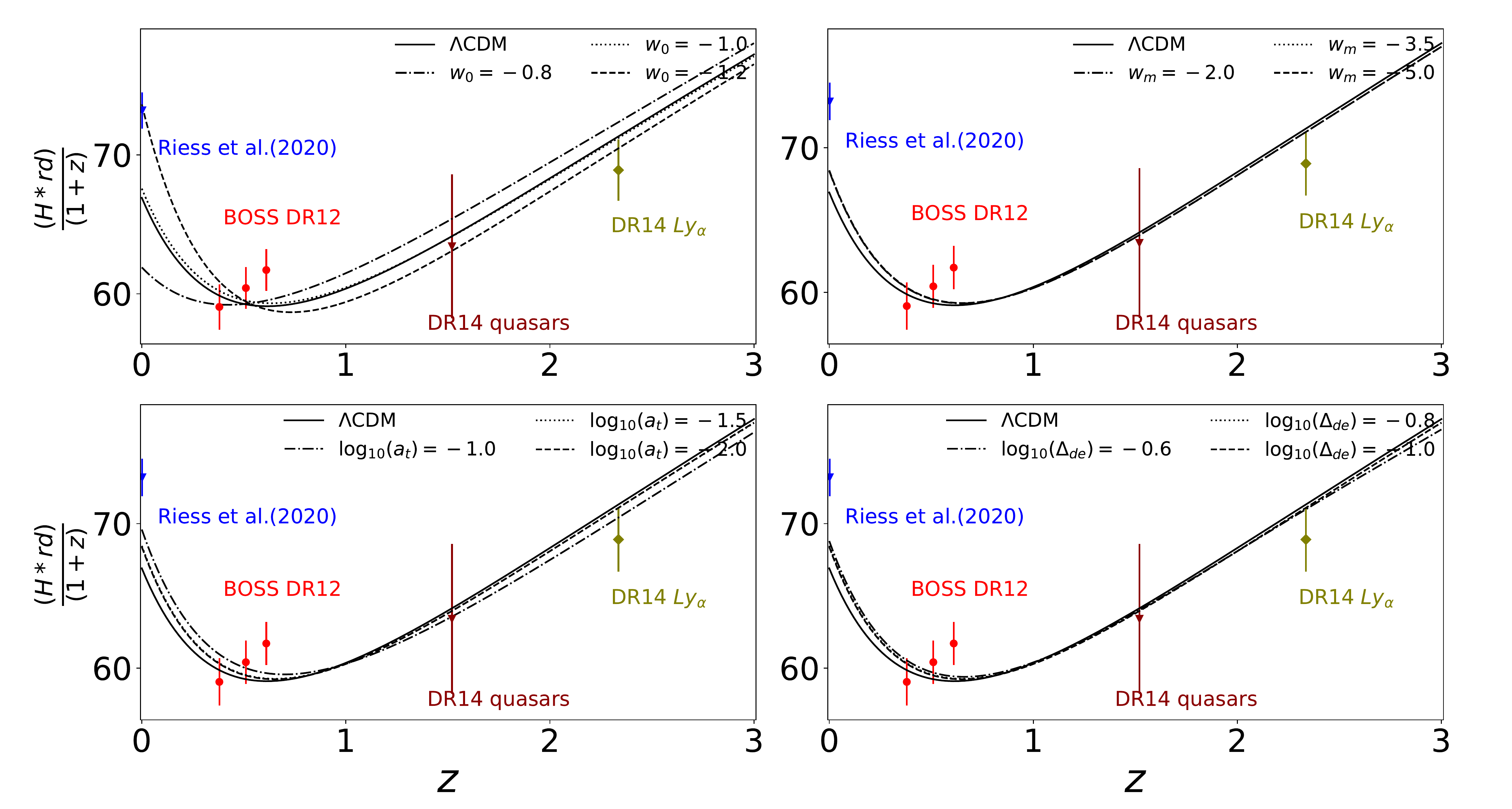}
    \caption{The plots of the evolution of $H(z) r_d/(1+z)$ for 4pDE models and reference $\Lambda$CDM model. Each of the four panels shows effect of varying  different dark energy equation of state parameter (see the legends) while fixing other EoS parameters to values $(w_0=-1.03,\  w_m=-3.97,\ \log_{10}(a_t)=-1.63,\ \log_{10}(\Delta_{\rm de})=-0.87)$.}
    \label{fig:4pde_hrd}
\end{figure}

\begin{figure}[hbt!]
    \centering
    \includegraphics[scale=0.38]{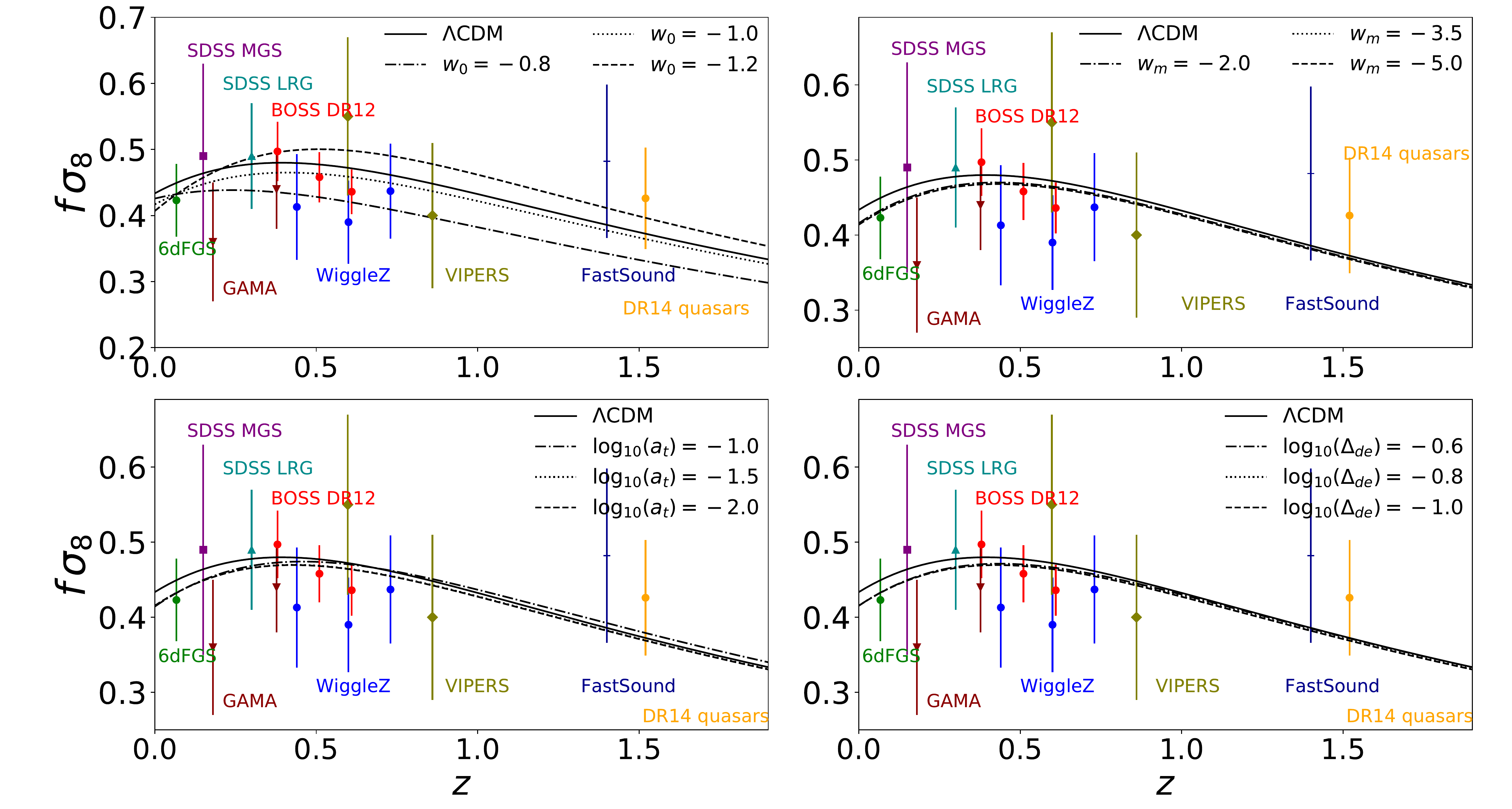}
    \caption{The plots of the evolution of the parameter $f\sigma_8$ as a function of redshift for 4pDE model and $\Lambda$CDM Model. Each of the four panels shows effect of varying  different dark energy equation of state parameter (see the legends) while fixing other EoS parameters to values $(w_0=-1.03,\  w_m=-3.97,\ \log_{10}(a_t)=-1.63,\ \log_{10}(\Delta_{\rm de})=-0.87)$.}
    \label{fig:4pde_fsig}
\end{figure}

\begin{figure}[hbt!]
    \centering
    \includegraphics[scale=0.38]{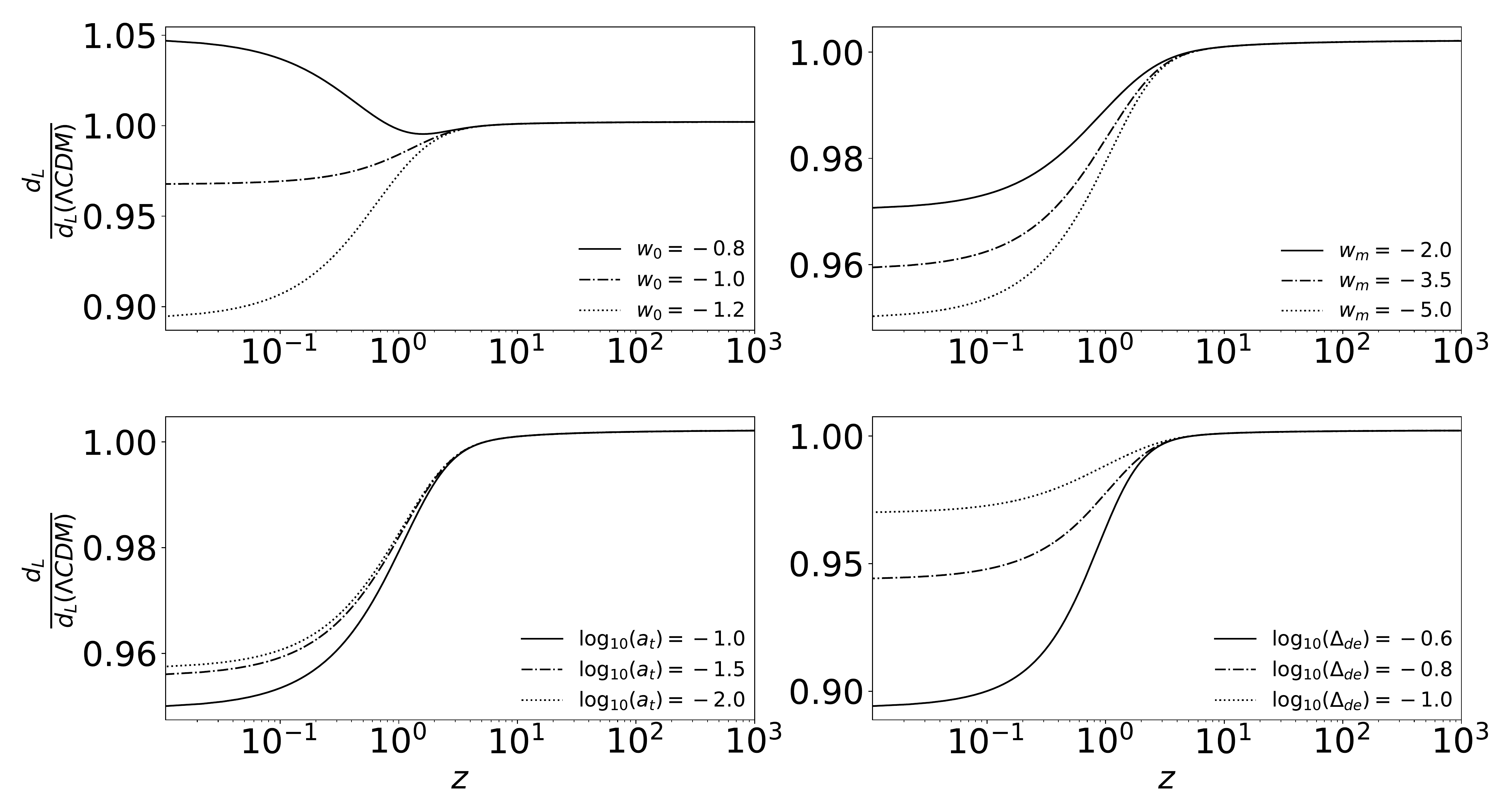}
    \caption{The plots of the evolution of the luminosity distance as a function of redshift for 4pDE model and $\Lambda$CDM model. Each of the four panels shows effect of varying  different dark energy equation of state parameter (see the legends) while fixing other EoS parameters to values $(w_0=-1.03,\  w_m=-3.97,\ \log_{10}(a_t)=-1.63,\ \log_{10}(\Delta_{\rm de})=-0.87)$.}
    \label{fig:4pde_dl}
\end{figure}

The dark energy equation of state parameter $w_0$ is most sensitive to observables because it is equation of state when the dark energy density is dominating the universe. So an increase in the value of this parameter rapidly increase the dark energy density and the expansion rate of universe hence leave a major impact on the late time observables. This fact is evident from upper left panel of Figures~\ref{fig:4pde_hrd},  \ref{fig:4pde_fsig} and \ref{fig:4pde_dl}.
The $w_m$ parameter is the DE equation of state at earlier times. At that time the amount of the dark energy density was relatively less. So this equation of state parameters has less impact on the observables.  The upper right panel of Figures~\ref{fig:4pde_hrd}, \ref{fig:4pde_fsig} and \ref{fig:4pde_dl} show the impact of varying $w_m$ on various observables. We can see that the change in the observables are relatively low in comparison to the case of parameter $w_0$.
The DE parameters $a_t$ and $\Delta_{\rm de}$ determine the width and the epoch of transition of the equation of state from $w_m$ to $w_0$. The effects of varying $a_t$ and $\Delta_{\rm de}$ are shown in the down-left and the down-right panels of the Figures~\ref{fig:4pde_hrd}, \ref{fig:4pde_fsig} and \ref{fig:4pde_dl} respectively. We can see that parameter $a_t$ has least impact on the observables given smooth (large width) transition preferred by the data. We also see that the DE parameter $\Delta_{\rm de}$ has comparatively more impact on observables.
 
\end{document}